\documentclass[12pt,preprint]{emulateapj}
\usepackage{amsmath,amssymb,threeparttable}

\slugcomment{Accepted to ApJ, September 6, 2013}

\shorttitle{Collisions in Debris Disks}
\shortauthors{Nesvold et al.}
\begin{document}

\title{SMACK: A New Algorithm for Modeling \\ Collisions and Dynamics of Planetesimals in Debris Disks}

\author{Erika R. Nesvold}
\affil{Department of Physics, University of Maryland Baltimore County
\\ 1000 Hilltop Circle
\\ Baltimore, MD 21250}
\email{Erika.Nesvold@umbc.edu}

\author{Marc J. Kuchner}
\affil{NASA Goddard Space Flight Center 
\\ Exoplanets and Stellar Astrophysics Laboratory, Code 667
\\ Greenbelt, MD 21230}
\email{Marc.Kuchner@nasa.gov}

\author{Hanno Rein}
\affil{Institute for Advanced Study 
\\ 1 Einstein Drive 
\\ Princeton, NJ 08540}
\affil{University of Toronto Scarborough
\\ 1265 Military Trail
\\ Toronto, Ontario M1C 1A4}
\email{rein@ias.edu}

\author{Margaret Pan}
\affil{NASA Goddard Space Flight Center
\\ Exoplanets and Stellar Astrophysics Laboratory, Code 667
\\ Greenbelt, MD 21230}
\email{Margaret.Pan@nasa.gov}

\begin{abstract}

We present the Superparticle Model/Algorithm for Collisions in Kuiper belts and debris disks (SMACK), a new method for simultaneously modeling, in 3-D, the collisional and dynamical evolution of planetesimals in a debris disk with planets. SMACK can simulate azimuthal asymmetries and how these asymmetries evolve over time. We show that SMACK is stable to numerical viscosity and numerical heating over $10^7$ yr, and that it can reproduce analytic models of disk evolution. We use SMACK to model the evolution of a debris ring containing a planet on an eccentric orbit. Differential precession creates a spiral structure as the ring evolves, but collisions subsequently break up the spiral, leaving a narrower eccentric ring.
\end{abstract}


\section{Introduction}
Spatially resolved debris disk images from optical and infrared observatories show spectacular patterns and sub-structures including eccentric rings \citep[e.g.][]{Kalas2005, Schneider2009, Krist2012, Boley2012}, warps or sub-disks \citep[e.g.][]{Golimowski2006, Krist2005}, and various other morphologies \citep[e.g.][]{Hines2007, Kalas2007}. Undetected exoplanets could create many of these features via gravitational perturbations. Many authors have analyzed resolved images of debris disks to predict the presence of exoplanets and constrain their locations, orbits, and physical properties \citep[e.g.][]{Wyatt1999a, Greaves2005, Quillen2006, Stark2008}. In the last few years, direct images of exoplanets combined with numerical models \citep[e.g.][]{Chiang2009, Lagrange2010} have demonstrated the power and necessity of this approach. 

However, our ability to use debris disk asymmetries as signposts of planets is limited by our ability to model debris collisions. The collisional lifetime of a dust grain with orbital period $t_{per}$ in a disk of optical depth $\tau_{eff}$ is given by $t_{col} = t_{per}/4 \pi \tau_{eff}$ \citep{Wyatt2008}. In many debris disk systems, this collisional timescale is shorter than the timescales for dynamical sculpting by planets, so collisions can have a significant effect on the dynamics of the disk compared to the gravity of the planet.

For example, \cite{Chiang2009} estimated the mass of a planet sculpting the Fomalhaut ring to be $M_{pl} \approx 0.5 M_J$ and the ring to have an optical depth of $\tau_{eff} \approx 10^{-3}$.  The secular dynamical perturbations from a planet of mass $M_{pl}$ act on a timescale of $t_{sec} \approx t_{per} M_*/M_{pl}$ \citep{Murray1999} but would have a collisional lifetime of $t_{col} \approx 3.8 \times 10^4$ yr, so a planetesimal orbiting Fomalhaut at 140 AU would experience secular effects on a timescale of $t_{sec} \approx 4 \times 10^6$ yr. Resonant effects, which could produce azimuthally asymmetric structures, can act on comparable timescales as well \citep{Kuchner2003}. In the Fomalhaut disk, the resonant timescale is $t_{res} \approx t_{per} (M_*/M_{pl})^{1/2} \approx 7 \times 10^4$ yr. Models of secular and resonant phenomena in such systems must therefore take collisions into account.

There have been many attempts to wed collisional and dynamical effects in a debris disk model. Early models adapted N-body simulations with routines that spawned new code bodies at every collision. For example, \cite{Beauge1990} represented the products of each fragmentation with five daughter bodies. \cite{Grigorieva2007} represented fragmentation products by introducing one daughter body per decade in mass. However, in this kind of algorithm, the number of code particles quickly increased to unmanageable numbers, so these models could not be run for very many orbits.

Other modelers took the opposite approach: they began with particle-in-a-box codes tracking averaged dynamical quantities and added spatial resolution and other refinements to increase the dynamical fidelity. For example, \cite{Kenyon2006} utilized a disk divided into a series of rings, each of which contained a particle-in-a-box calculation. The ACE code \citep{Krivov2005,Lohne2008,Vitense2012} utilizes the Boltzmann-Schmoluchowski equation to model collisional and dynamical evolution in debris disks with dynamics averaged over the angular orbital elements. These kinds of codes can simulate $10^9$ yr of disk evolution on existing computers \citep[e.g.][]{Vitense2012}, but cannot be used to model disk asymmetries in three dimensions.

Still other methods employ fully three-dimensional dynamics, but only aim to model systems that are in steady state, i.e., the sources and sinks of dust grains roughly balance one another, as they might in a system that has already evolved for many collision times. One method that makes this approximation is the collisional grooming algorithm \citep[e.g.][]{Stark2009, Kuchner2010}. Another is DyCoSS \citep{Thebault2012}. 

A new model, LIPAD \citep{Levison2012}, uses a superparticle method coupled with full N-body integrators to simulate collisional and dynamical evolution, while keeping the number of superparticles roughly constant. Using this approach avoids the steady-state approximation. We will discuss superparticle methods in Section \ref{sec:superparticles}.

But despite all this work, no published collisional/dynamic disk models quite met our desires for interpreting images of debris disks. We wanted a model that could
\begin{itemize}
\item Track collisional and dynamical evolution of debris disks in 3-D to model asymmetries created by planets like warps and eccentric rings, and
\item Run stably for $10^7$ or more years of disk evolution in a feasible number of CPU cycles.
\end{itemize}
The above algorithms did not meet these requirements. For example, the LIPAD code uses a scale-height approximation for the disk's vertical structure, and the longest published LIPAD simulations ran for only $10^4$ yr.

Therefore, we have developed a new tool for 3-D modeling of collisional planetesimal populations in debris disks. Our tool, the Superparticle-Method Algorithm for Collisions in Kuiper belts and debris disks (SMACK), uses a superparticle approximation to simultaneously track the N-body dynamics and collisional evolution of the bodies that produce the dust we observe and calculate the dust production rates. We have designed SMACK with the ultimate goal of deriving improved estimates of the masses and orbital parameters of exoplanets in debris disks using high spatial resolution images, e.g., from the Atacama Large Millimeter Array (ALMA). This paper describes the SMACK algorithm, and presents a basic SMACK model for an eccentric debris ring.

\section{The Modeling Tool}
\label{sec:methods}

\subsection{Superparticles}
\label{sec:superparticles}

Our full numerical model uses an N-body integrator, REBOUND, to solve the equations of motion of the planetesimals and detect collisions, combined with a collision resolution algorithm, SMACK, to calculate the effects of collisions on the velocities and size distributions of the planetesimals. REBOUND \citep{Rein2012} is a multi-purpose code originally designed for studying collisional dynamics in planetary rings, freely available under an open-source license from \url{https://github.com/hannorein/rebound}. While the \cite{Grigorieva2007} model and LIPAD both detect collisions using a two-dimensional grid, REBOUND detects collisions in 3-D, without using a grid; it contains a Barnes-Hut tree module to calculate self-gravity and detect collisions, parallelized with MPI and OpenMP. 

In the standard version of REBOUND, collisions are resolved using an instantaneous collision model with a normal coefficient of restitution. This approximation handles a variety of problems in planetary ring dynamics. Debris disks contain a very different physical regime than planetary rings; collision speeds in debris disks are much higher than in rings (km $\hbox{s}^{-1}$ vs. mm $\hbox{s}^{-1}$) and optical depths are lower ($\sim 10^{-4}$ vs $\sim 1$). So rather than allowing each REBOUND code body to represent one planetesimal, we use each body as either a planet with mass, or as a massless ``superparticle'' representing a group of planetesimals on similar trajectories. 

Superparticle methods have already been used extensively to model planetesimal formation within gas disks \citep[e.g][]{Michikoshi2009, Rein2010, Zsom2008, Johansen2012, Charnoz2012}. They have also been applied to model planetesimals in the solar system and in debris disks. \cite{Charnoz2003} used a superparticle approach to study how the dynamics of particles ejected from the Jupiter-Saturn region affected their size distributions, though they did not include the feedback from the collisions on the particle dynamics. \cite{Grigorieva2007} used cylindrical superparticles ~5 AU in diameter fixed to the disk midplane to model collisional avalanches in debris disks over spans of $\sim 40$ orbital periods. The LIPAD model \citep{Levison2012} uses a superparticle approach; these authors refer to their superparticles as ``tracers''. 

\subsection{SMACK}
\label{sec:smack}

In SMACK, the superparticles represent collections of planetesimals with a range of sizes, as in \cite{Charnoz2003}. However, in \cite{Charnoz2003}, the evolution of the size distribution of each superparticle is calculated only after the entire dynamical integration of the superparticles is complete. In contrast, SMACK calculates the size evolution at each timestep of the N-body integration, allowing us to model the feedback between the collisions and the dynamics of the superparticles. Each superparticle in SMACK is characterized by an incremental size distribution with logarithmic size bins.

In general, two colliding, fragmenting bodies produce a spray of daughter particles with different trajectories. Simulating this distribution of trajectories in great detail would require SMACK to create new superparticles with each collision, quickly increasing the number of bodies tracked by REBOUND with every collision. Instead, SMACK approximates the outcomes of collisions while keeping the number of integrator bodies constant. When REBOUND detects an encounter between two superparticles, it passes the velocities and size distributions of the overlapping superparticles to SMACK. SMACK returns two superparticles with different velocities and size distributions and REBOUND continues its dynamical integrations using these modified superparticles. 

The essence of SMACK is simple. In SMACK, the fragments produced by collisions are swapped between superparticles, so that the new size distributions $n_a$ and $n_b$ are given by 
\begin{equation}\label{eq:na} n_a(i) = n_A(i) - P_A(i) + F_B(i) \end{equation}
\begin{equation}\label{eq:nb} n_b(i) = n_B(i) - P_B(i) + F_A(i), \end{equation}
where $P_A(i)$ is the number of parent body particles in size bin $i$ in superparticle $A$ that are lost due to collisions, $F_B(i)$ is the number of daughter particles in size bin $i$ produced by colliding particles in superparticle $B$, and $n_A$ and $n_B$ are the size distributions of the parent superparticles. The detailed forms of $P(i)$ and $F(i)$ used in SMACK are given in Section \ref{sec:fragmentation}.

This swapping of fragments is a first-order approximation of the velocity distribution of fragments in a planetesimal collision. In a collision between two real planetesimals, the fragments are produced in roughly the center-of-mass frame. But an encounter between two superparticles is more complicated; the center of mass of the superparticles is not the same, in general, as the center of mass of any pair of planetesimals represented by the superparticles. If the mass ratio of the parent bodies is higher (lower) than that of the superparticles, the fragments should be launched in a direction skewed toward that of the more (less) massive parent bodies. The swapping described above crudely approximates this physics.

After swapping the daughter planetesimals, SMACK calculates the kinetic energy lost in the collisions and corrects the superparticle velocities to reflect this loss and to conserve momentum. Let $v_A$ and $v_B$ be the magnitudes of the velocities of the parent superparticles in the center-of-momentum frame and $m_A$ and $m_B$ be the total masses of the parent superparticles. Some fraction of the kinetic energies of the parent superparticles will be lost to planetesimal collisions. This fraction depends on the fraction of planetesimals that experience collisions and the amount of kinetic energy lost by each planetesimal in a collision. Using the energy loss fraction and the collision rates for each pair of size bins, SMACK calculates $E_A$ and $E_B$, the kinetic energy lost to collisions in superparticles A and B, respectively (as described in Section \ref{sec:fragmentation}). Then the new superparticle velocities must satisfy the energy conservation law,
\begin{equation}\label{eq:energy}K = \frac{1}{2} m_a v_a^2 + \frac{1}{2} m_b v_b^2 = \frac{1}{2} m_A v_A^2 + \frac{1}{2} m_B v_B^2 - E_A - E_B, \end{equation} 
where $K$ is the total kinetic energy of the two superparticles, and $m_a$ and $m_b$ are the new total masses of each superparticle, calculated from the new post-encounter size distributions given by equations (\ref{eq:na}) and (\ref{eq:nb}). The velocities must also satisfy the momentum conservation law,
\begin{equation}\label{eq:momentum} m_a v_a + m_b v_b = 0 \end{equation}
The velocities that solve Equations (\ref{eq:energy}) and (\ref{eq:momentum}) are
\begin{align} \label{eq:va} v_a &= \sqrt{\frac{2 m_b K}{m_a (m_a + m_b)}}
\\ \label{eq:vb} v_b &= - \frac{m_a}{m_b} v_a.
\end{align}
Equations (\ref{eq:na}), (\ref{eq:nb}), (\ref{eq:va}), and (\ref{eq:vb}) give the new size distributions and velocities of the superparticles, and define the essential SMACK algorithm.

When a superparticle encounter yields no planetesimal collisions (i.e., all the planetesimals pass through the encounter unaffected), the algorithm gives exact results. When all the planetesimals in each superparticle collide, the algorithm produces an outcome that is a good approximation for the dominant size bins. When there is a mix of collisions and pass-throughs, the algorithm compromises, making errors in the distribution of output velocities, but not in the total energy or angular momentum.

\subsection{Fragmentation}
\label{sec:fragmentation}

For now, SMACK only models one type of collision outcomes: catastrophic collisions, defined as collisions in which the largest fragment is no larger than half the size of the target. Cratering collisions do not have a significant effect on the steady-state size distribution of a collisional cascade \citep{Dohnanyi1969}, so we neglect them for now. Since we are modeling disks in which the impact velocities are high, we also ignore bouncing, sticking, and gravity between planetesimals. Future versions of SMACK may incorporate these effects.

Consider two superparticles, A and B, that are found to overlap during a given timestep. The number of planetesimals from size bin $i$ in superparticle A that collide with planetesimals in size bin $j$ in superparticle B is
\begin{equation}\label{eq:numcol} c_A(i,j) = n_A(i) \tau_B(i,j),\end{equation}
where $n_A(i)$ is the number of planetesimals in size bin $i$ in superparticle A and $\tau_B(i,j)$ is the optical depth along the path of a planetesimal in size bin $i$ passing through superparticle B for collisions with planetesimals of size $j$. SMACK estimates this optical depth as
\begin{equation}\label{eq:od} \tau_B(i,j) \approx n_B (j) \sigma_{ij} l/V, \end{equation}
where $\sigma_{ij}$ is the combined collisional cross-section of a planetesimal in size bin $i$ and a planetesimal in size bin $j$, $l$ is the path length traveled by a planetesimal in superparticle A, and $V$ is the volume of superparticle B. The path length $l$ is the distance traveled by superparticle A through the disk relative to the local Keplerian flow since its last encounter. SMACK estimates this distance as $l = v_{AB} t_{enc}$, where $v_{AB}$ is the magnitude of the relative velocity of superparticles A and B, and $t_{enc}$ is the time since superparticle A's last encounter. The superparticle volume $V$ is the parameter used by REBOUND to determine when a collision occurs. The superparticle volume must be chosen carefully to minimize both computation time (see Section \ref{sec:smallloop}) and numerical heating (see Section \ref{sec:heating}). We run numerical heating tests before every new simulation to select the optimal superparticle size. The collisional cross-section is purely geometric because gravitational effects between planetesimals are ignored. The cross-section is given by
\begin{equation}\label{eq:crosssection} \sigma_{ij} = \frac{\pi}{4} (D(i) + D(j))^2, \end{equation}
where $D(i)$ and $D(j)$ are the diameters of planetesimals in size bins $i$ and $j$, respectively.

We aspire to model a system in which each planetesimal involved in a fragmenting collision loses some fraction $f_{KE}$ of its kinetic energy. This fraction $f_{KE}$ is a parameter of the code. The total desired energy loss for size bin $i$ in superparticle $A$ during its encounter with superparticle $B$ is calculated by SMACK using
\begin{equation}\label{eq:energyloss} E_A(i) =  \displaystyle\sum\limits_{j} f_{KE} \frac{1}{2} m(i) v_A(i,j)^2 c_A(i,j), \end{equation}
where $v_A(i,j)$ is the magnitude of the velocity of superparticle $A$ in the center-of-momentum frame of a planetesimal in size bin $i$ in $A$ and a planetesimal in size bin $j$ in $B$.

Some of the colliding planetesimals counted in equation (\ref{eq:numcol}) may shatter, creating a distribution of smaller fragments. SMACK determines which colliding planetesimals will fragment by comparing the collision energy of each pair of colliding planetesimals to the minimum kinetic energy needed for a catastrophic collision. In the center-of-mass frame, the kinetic energy of two colliding planetesimals with masses $m(i)$ and $m(j)$ is
\begin{equation} E_{col} = \frac{1}{2} \frac{m(i) m(j)}{m(i) + m(j)} v_{rel}^2, \end{equation}
where $v_{rel}$ is the magnitude of the relative velocity of the two planetesimals. 
In laboratory experiments, \cite{Hartmann1980} found that approximately half of this collisional kinetic energy is partitioned into each colliding body, regardless of the mass ratio of the two, so a planetesimal in size bin $i$ will shatter in a collision with a planetesimal in size bin $j$ if 
\begin{equation}\label{eq:condition} \frac{1}{2} E_{col}(i,j,v_{rel}) \geq E_{min}(i), \end{equation}
where $E_{min}(i)$ is the minimum energy needed to shatter a planetesimal in size bin $i$.

The minimum shattering energy, $E_{min}(i)$, is a combination of the gravitational binding energy of the planetesimal and its internal impact strength. We use the minimum energy criteria derived by \cite{Durda1993}, 
\begin{equation} E_{min}(i) = \frac{1}{f_{KE}} \left( 0.822 \frac{G m(i)^2}{D(i)} + \frac{1}{6} \pi S D(i)^3\right), \end{equation}
where $G$ is the gravitational constant, $m(i)$ is the mass of planetesimals in size bin $i$, $D(i)$ is the diameter of planetesimals in size bin $i$, and $S$ is the impact strength of the planetesimals. We use $f_{KE} = 0.1$ \citep{Fujiwara1982} and a size-independent impact strength of $S = 3 \times 10^6 J\cdot m^{-3}$ \citep{Greenberg1977}.

If Equation (\ref{eq:condition}) holds true, the loss of parent bodies in size bin $i$ due to fragmentation in collisions with size bin $j$ is equal to the number of collisions between planetesimals in $i$ and $j$:
\begin{equation} P_A(i,j) =c_A(i,j). \end{equation}
If the inequality in Equation (\ref{eq:condition}) does not hold, the planetesimal in size bin $i$ will not shatter and will not be counted as loss, so $P_A(i,j) = 0$. The planetesimal in size bin $j$ may fragment, however, depending on whether $E_{min}(j)$ satisfies Equation (\ref{eq:condition}); SMACK performs these calculations by looping through one index at a time. The total loss in size bin $i$ in superparticle $A$ is the sum of the losses due to each size bin in superparticle B:
\begin{equation}\label{eq:lossAtot} P_A(i) = \displaystyle\sum\limits_{j} P_A(i,j) . \end{equation}

Collisions produce fragments in power law size distributions, which we represent with incremental logarithmic bins. The individual daughter particle distribution resulting from fragmentation in size bin $i$ is
\begin{equation} \label{eq:fragments} F_A(i,j) = P_A(i,j) \kappa_AD(i)^{\alpha} . \end{equation}
The fragment distribution index, $\alpha$, is a parameter of the algorithm. The constant $\kappa_A$ is calculated for every collision for each superparticle such that the largest fragment in each distribution is one-half the mass of the parent planetesimal that produces the fragments. Again, we find the total fragment gain from size bin $i$ in superparticle A by summing over the size bins in superparticle $B$ that collide with $i$:
\begin{equation} F_A(i) = \displaystyle\sum\limits_{j} F_A(i,j). \end{equation}

\subsection{The Smallest Planetesimals}
\label{sec:smallloop}

If the total optical depth for a given size bin is ever greater than one, i.e.,
\begin{equation} \tau_B(i) =  \displaystyle\sum\limits_{j} \tau_B(i,j) > 1,\end{equation}
the loss in that size bin given by equation (\ref{eq:lossAtot}) could be greater than the number of planetesimals in that bin. We generally try to avoid this situation by choosing the number and volume of the superparticles so that the superparticle encounter time is less than the collision time for the small grains. However, it sometimes occurs anyway as the disk evolves and more small planetesimals are produced.  This situation tends to arise for the smallest planetesimals first, since they collide most frequently, and become more common with increasing superparticle encounter times and with increasing optical depths. 

To address this issue, we adopted a variable-timestep method. If the maximum optical depth for all the size bins in a superparticle is ever found to be greater than 1, SMACK divides the path length traveled by the superparticle since its last encounter into segments such that the maximum optical depth in any size bin is $\leq 1$. After each segment, the optical depths are recalculated and the energy and planetesimal losses are calculated as described above. The relative velocity of the superparticles is updated after each segment to accurately reflect the energy lost to collisions, but the superparticle trajectories are not changed until the final segment. 

This ``small particle loop'' allows us to run SMACK uninterrupted in regions of unexpectedly high density with fewer superparticles. However, it also introduces noise into the velocity evolution of the superparticles, as the trajectories are not updated during the loop. We therefore check each astrophysical simulation by running it with a few different superparticle sizes to ensure that we get the same result, and that the small particle loop is not introducing excess noise.

While REBOUND can be adapted to model small dust grains by adding addition forces to the integrator such as radiation pressure and Poynting-Robinson drag, SMACK cannot simultaneously model dust grains and planetesimals within the same superparticles if they are subject to different forces. We therefore model only grains larger than 1 mm, which do not experience significant radiative forces during their collisional lifetimes, even in a very sparse disk like the zodiacal cloud. The current version of SMACK thus cannot directly simulate disk images at optical wavelengths, but is meant for simulating high-resolution millimeter and sub-millimeter images from instruments like ALMA.

\subsection{Normalization}

To compare our models with images and photometry of known debris disks, we need to know the face-on optical depth, $\tau_{disk}$, of the models, and how it relates to $\tau_{SP}$, the total cross section of the planetesimals in the superparticle divided by the cross section of the superparticle, $\pi r_s^2$. Equation (\ref{eq:od}) above implies that the density of an individual superparticle models the local disk density. So $\tau_{disk}$ is related to the individual optical depth of each superparticle by a linear filling factor, $f_{SP}$, where
\begin{equation} \tau_{disk} = f_{SP} \tau_{SP}. \end{equation}
The filling factor, $f_{SP}$, is simply the average number of superparticles that a perpendicular path through the disk would intersect. In other words, $f_{SP}$ is the inverse of the fraction of the perpendicular path through the disk that would be contained within one superparticle. This factor is approximately 
\begin{equation} \label{eq:fsp} f_{SP} \approx \frac{3h}{4r_s}, \end{equation}
where $h$ is the full height of the planetesimal distribution and $4r_s/3$ is the average length of a chord through a spherical superparticle with radius $r_s$.

We calculate $f_{SP}$ before the simulation begins using equation (\ref{eq:fsp}). We generate $10^6$ superparticles with the same orbital element distributions that we will use as the initial conditions. Then we choose a fiducial perpendicular path through the disk, and create a histogram of the positions of the superparticles along that path. We estimate $h$ along that path by normalizing the histogram such that the maximum value is $1$ and summing together all the bins. Knowing $f_{SP}$ allows us to set the total cross section of the planetesimals in the superparticles to yield an initial face-on optical depth of our choosing. 

The planetesimal size distributions in the superparticles begin at $1$ mm, but dust grains in disks are ground down to $\sim 1$ $\mu$m in size before they are removed from the system by radiation pressure. To ensure that we are including the contribution of these smaller dust grains to the cross-sectional area of the planetesimals in the disk, we fit a power law to the size distributions of each superparticle, then extrapolate the size distributions down to $1$ $\mu$m. We use the extrapolated size distributions when calculating the cross-sectional area of the superparticles during normalization.

\section{Numerical Tests}
\label{sec:tests}

We performed a series of numerical tests on SMACK to validate the algorithm and identify sources of numerical noise. We used the Wisdom-Holman integrator \citep{Wisdom1991} included in REBOUND, which closely follows the implementation of the SWIFT code \citep{Levison1994}. For collision detection, we selected REBOUND's tree algorithm, which implements a nearest-neighbor search to find overlapping superparticles at each timestep. We ran REBOUND and SMACK on the NASA Center for Climate Simulation's (NCCS) Discover cluster, using a hybrid OpenMP/MPI parallelization on 48 cores.

In each simulation, we assume the planetesimals are spherical with a density of 3 $g/cm^3$. We also assume that the power law distribution of the fragments (see equation \ref{eq:fragments}) has index $\alpha = -2.8$. We also assume the star has mass 1 $M_\odot$ and radius 1 $R_\odot$. REBOUND outputs the orbital elements, Cartesian coordinates, and size distributions of each superparticle every output timestep, where the size of the output timestep is greater than the integrator timestep and is set by the user. For each simulation, we use an integrator timestep of 1 yr and an output timestep such that REBOUND outputs 1000 total data points for each superparticle. For example, we use the output timestep of $10^4$ yr for a $10^7$ yr simulation.  We use open boundary conditions for our systems; if at any timestep a superparticle's orbital elements place it outside a cube with a user-defined width $l_{box}$ centered on the star, the superparticle is removed from the simulation. The system boundaries form a cube because REBOUND was originally developed for studying shearing boxes.  If a superparticle collides with the star or any planet in the system, the superparticle is also removed. The initial conditions for each simulation presented in this section are shown in Table (\ref{tab:initial}). The angular orbital elements (longitude of the ascending node $\Omega$, argument of pericenter $\omega$, and true anomaly $f$), are all distributed uniformly between $0$ and $2\pi$ for each simulation.

\subsection{Size Distribution Evolution}
\label{sec:sizedist}

\cite{Dohnanyi1969} studied collisional cascades analytically assuming an infinite range of particle masses with each body experiencing a constant impact velocity, assuming a mass-independent material strength. He found that the incremental mass distribution of such a collisional cascade at steady state can be described by a power law with index $q = -1.833$ (with linear mass bins). The corresponding incremental size distribution with logarithmic bins is a power law with index $p = -2.5$ \citep{Durda1993}. 

\cite{Dohnanyi1969} also found that the index of the equilibrium size distribution is independent of the fragment size distribution index $\alpha$. \cite{Durda1993} confirmed this in his collisional model, and noted that steeper values of $\alpha$ corresponded to a faster convergence of the size distribution to an equilibrium power law. We chose a relatively steep value of $\alpha =-2.8$ to decrease the computation time required for the size distribution evolution tests. 

As an initial test of the collisional evolution simulated by SMACK, we placed 1000 superparticles around a solar-mass star with semi-major axes uniformly distributed between 90 AU and 110 AU. The superparticles were given eccentricities uniformly distributed between 0 and $e_{max} = 0.2$ and inclinations between 0 and $i_{max}=e_{max}/2$. The other orbital elements were uniformly distributed between 0 and $2\pi$. We distributed the planetesimals in each superparticle among 31 logarithmic size bins ranging from 1 mm in diameter to 1 m using a logarithmic increment of 0.1. In order to keep the average impact velocity constant to match the \cite{Dohnanyi1969} scenario, we turned off the velocity corrections in SMACK by allowing SMACK to update the size distribution of each superparticle after an encounter without updating the superparticle trajectories.

If the assumption of an infinite collisional cascade is relaxed and a cutoff is present in the small-sized end of the size distribution, a wave-like pattern emerges in the size distribution \citep{CampoBagatin1994}. A real cutoff of this sort occurs in disks at the particle blowout size for the system, producing real wave patterns, but artificial wave patterns can also appear as numerical noise in collisional models with artificial particle size cutoffs. We encountered this wave in our initial tests of the collisional algorithm in SMACK. Our simulated disk had an artificial cutoff in the size distribution at 1 mm. 

To remove this artificial wave, we use the method of \cite{Durda1993}. For each superparticle encounter, SMACK extrapolates the size distribution for each superparticle to 30 smaller size bins down to 1 $\mu$m. The number of collisions due to planetesimals in each of these smaller bins is calculated using equation (\ref{eq:numcol}) and is included in the particle loss and energy loss calculations (equations \ref{eq:lossAtot} and \ref{eq:energyloss}) for each superparticle. This small-particle extrapolation reduces the wave effect of the cutoff to negligible levels.

\begin{figure}[!ht]
	\centering
	\includegraphics[scale=0.4]{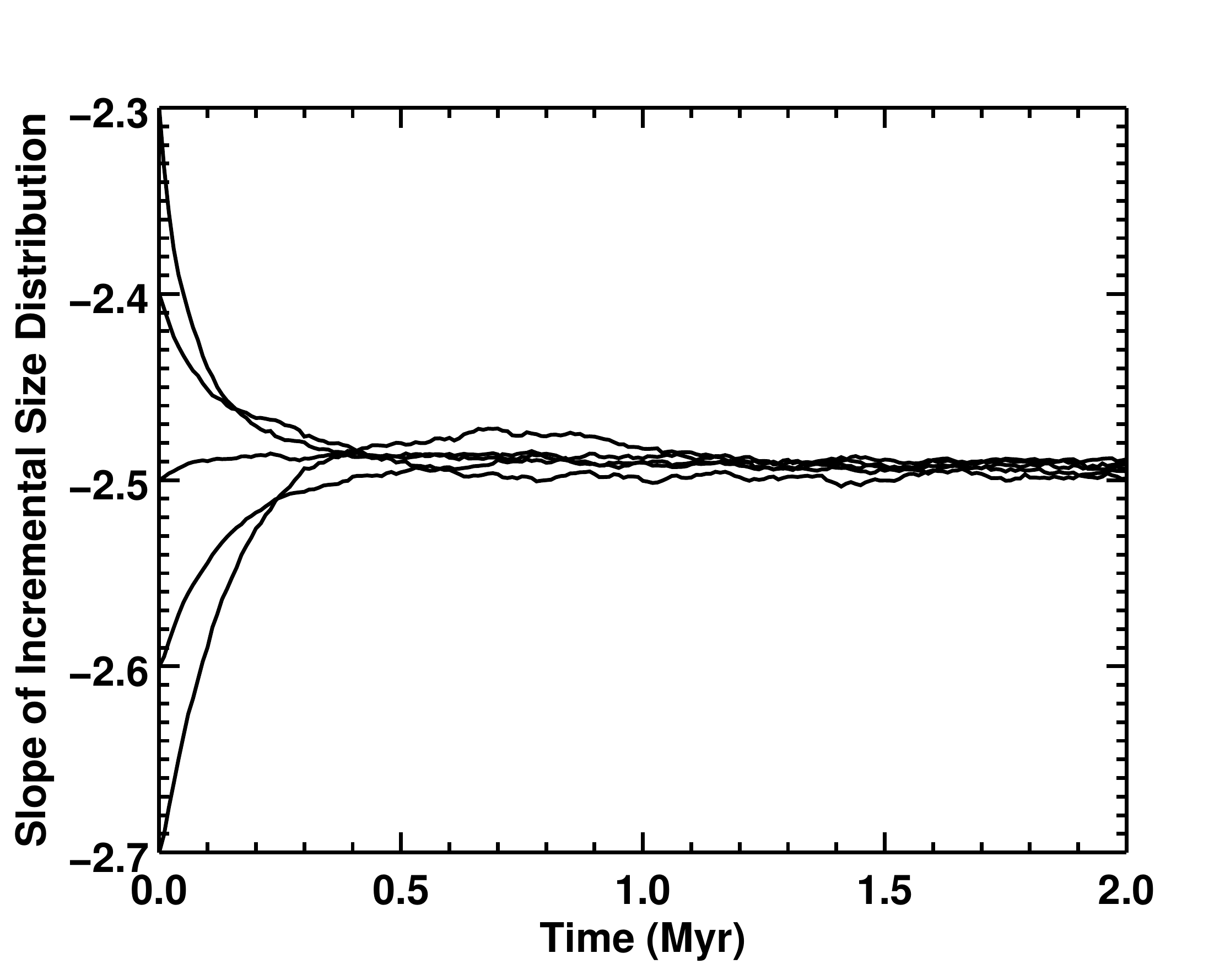}
	\caption{Incremental size distribution of the planetesimals in each of five different SMACK simulations. Each simulation had an initial power law size distribution of planetesimals between $1$ mm and $1$ m, with power law indices ranging from $p_0 = -2.3$ to $p_0 = -2.7$. Each size distribution equillibrated to a power law with index $p \approx -2.5$ within $2 \times10^6$ yr \citep{Dohnanyi1969}.} 
	\label{fig:slopes}
\end{figure}

We tested SMACK's ability to reproduce the convergence to a \cite{Dohnanyi1969} size distribution by running five simulations with different initial indices for the total size distributions for the planetesimals in the disk. For each simulation, set the vertical optical depth of the disk to $10^{-2}$. We summed the size distribution over all superparticles and fit a power law to the total size distribution at each output timestep. Then we compared the evolution of the power law indices over time for each of the five simulations. The results are shown in Fig. (\ref{fig:slopes}). As each collisional cascade evolved, the size distribution index grew or shrank until it reached the \cite{Dohnanyi1969} equilibrium index of $-2.5$. The size distribution index for each simulation converged to the \cite{Dohnanyi1969} index within $2 \times 10^6$ yr. For comparison, the collision time for the smallest planetesimals is $\sim 10^5$ yr in this simulation.

Using a size-dependent breaking strength \citep[e.g.][]{Gaspar2012a} or a size-dependent velocity distribution \citep[see][]{Pan2011} can cause breaks in the steady-state distribution of a collisional cascade. We did not attempt to reproduce these phenomena in these initial numerical tests.

\subsection{Numerical Heating}
\label{sec:heating}

Simulations of bouncing collisions suffer from numerical heating associated with the finite size of the superparticles. The superparticles in SMACK do not simply bounce, but the simulations of bouncing collisions can nonetheless inform us about the numerical noise in SMACK. For example, \cite{Lithwick2007} found that their grid-based simulations of bouncing collisions contained numerical heating on the scale of their grid size. Likewise, we expect that SMACK cannot model disks where the mean eccentricity of the planetesimals is $< r_s/r$ or the mean inclination of the planetesimals is $< r_s/r$ where $r$ is the radius of the disk.

We studied the eccentricity evolution of SMACK simulations as a function of superparticle size to search for additional numerical heating effects, like any artificial viscous stirring associated with the finite superparticle size \citep[see][]{Goldreich1978}. We simulationed a planetless ring with radius 90-110 AU and optical depth $5 \times 10^{-3}$ around a solar-mass star. We varied the superparticle radius between runs while varying the number of superparticles to keep the superparticle encounter time constant and measured how the mean eccentricity of the planetesimals evolved. We utilized the results of this test to choose the maximum superparticle size to use for the simulation described in Section \ref{sec:application}. 

At first, we found that simulations using a large range of planetesimal sizes (1 mm - 1 m) cooled more slowly, and the eccentricity evolution curves never converged for any value of $r_s$. This situation probably resulted from the coupling between planetesimals of various sizes. To reduce this kind of numerical noise, we decided to restrict the size range of the planetesimals in each superparticle to 1 mm - 10 cm for all simulations that included velocity evolution.

\begin{figure}[!ht]
	\centering
	\includegraphics[scale=0.4]{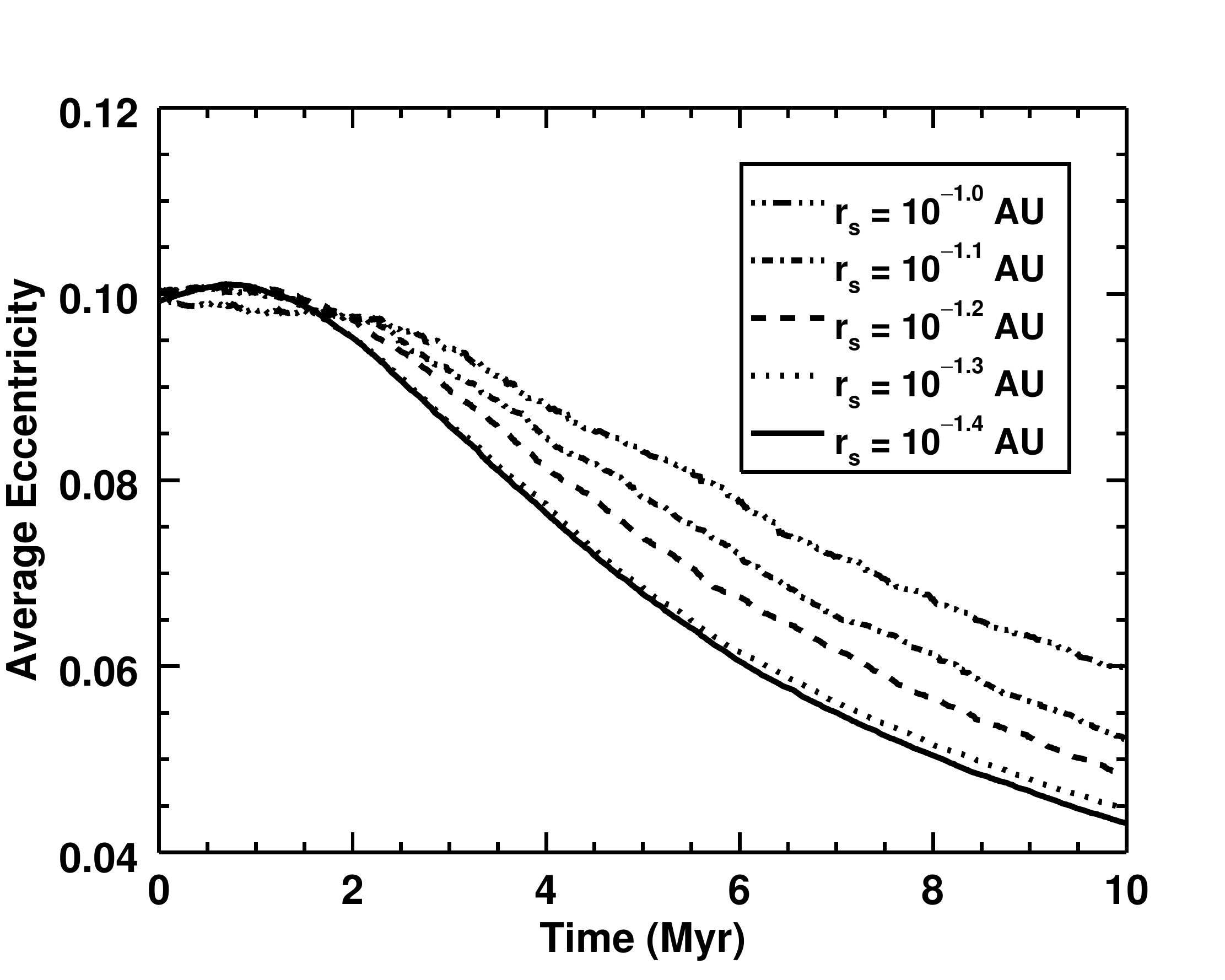}
	\caption{The average eccentricity of the planetesimals in a ring with no planet, with radius 90-110 AU, over 10 Myr for a series of simulations that model the same disk using different superparticle sizes. The damping rates converge for for superparticle radii $\approx 10^{-1.3}$ AU, suggesting that numerical heating is negligible for superparticles of this size.}
	\label{fig:heating}	
\end{figure}

Fig. (\ref{fig:heating}) shows the simulations with the reduced range of planetesimal sizes. The mean eccentricity of the superparticles in each simulation decreased rapidly at first due to inelastic collisions, then flattened and leveled off at a nonzero eccentricity. As we decreased the superparticle radius, the eccentricity evolution curved converged to a single damping curve, in which the mean eccentricity dropped by a factor of $1/2$ in $0.53$ Myr, or approximately $2\tau_{col}$, where $\tau_{col}$ is the collision timescale for the smallest planetesimals. The damping curves converged at a superparticle radius of $r_s = 10^{-1.3}$ AU, showing that this superparticle size probably suffices to mitigate numerical heating in our astrophysical simulations.

\subsection{Numerical Viscosity}
\label{sec:spreading}

\begin{figure}[!ht]
	\centering
	\includegraphics[scale=0.4]{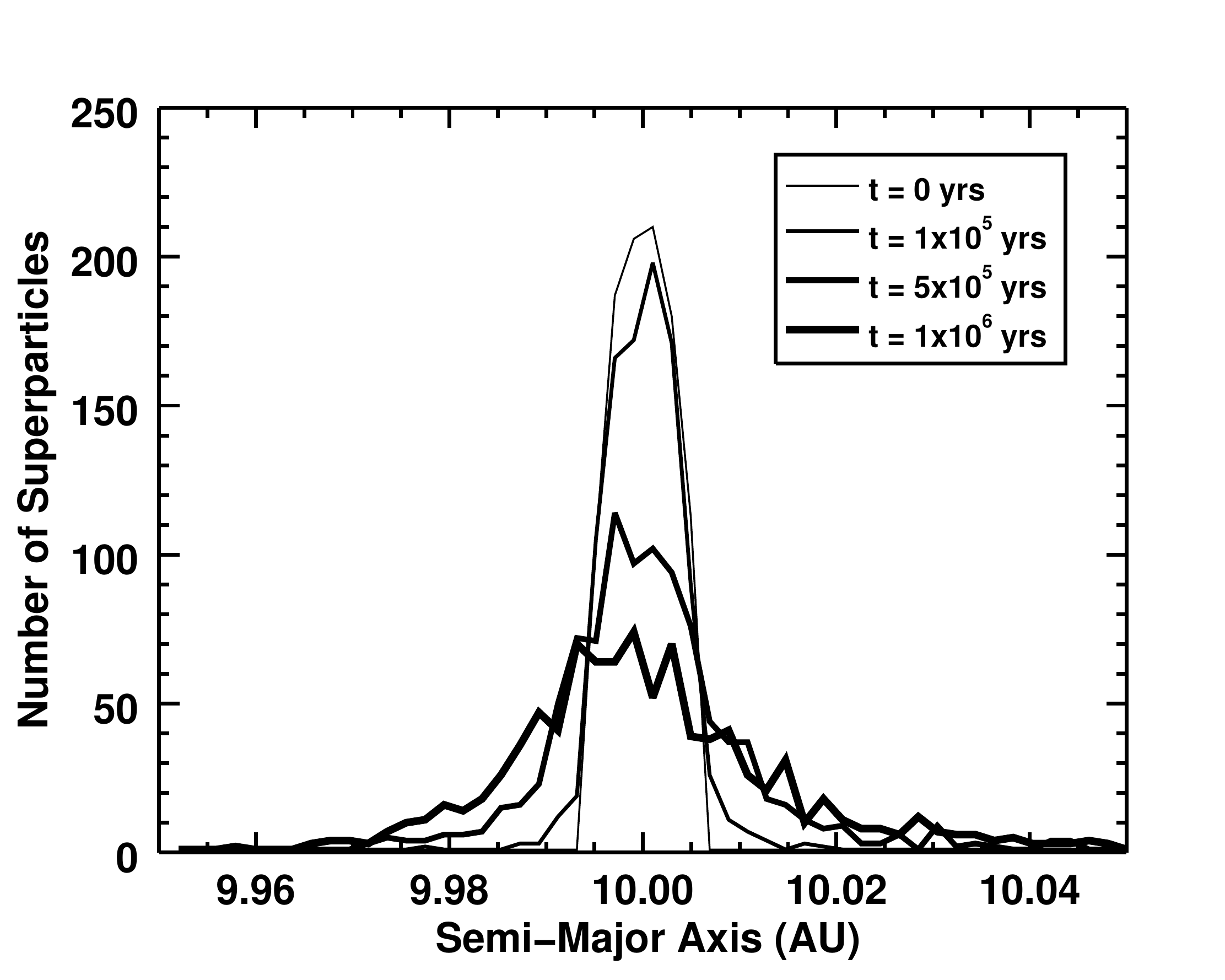}
	\caption{Histograms of superparticle semi-major axis at logarithmic time increments of a planetless ring at 10 AU with initial width 0.01 AU and superparticle radius $10^{-2.5}$ AU. Numerical viscosity causes this extremely narrow ring to spread by a factor of two in $6.67\times10^5$ yr.}
	\label{fig:spreading}
\end{figure}

The finite size of the superparticles also leads to diffusion of angular momentum, which can cause a narrow ring to spread on an artificially short timescale \citep[e.g.][]{Lithwick2007}. We tested our models for numerical viscous spreading by simulating the evolution of a very narrow ($\Delta r = 0.01$ AU) planet-less ring at $r = 10$ AU with superparticle radii of $10^{-2.5}$ AU for 1 Myr. Fig. (\ref{fig:spreading}) shows the histogram of the number density of superparticles at several times during the simulation. The spreading time of this ring, the time it takes for the width of the ring to double, is $\tau_{spread} = 6.67\times10^5$ yr. 

\cite{Petit1987} solved the diffusion equation for a collisional ring and found that the width of the ring $\Delta r$ at time $t$ is related to the particle size (in our case, the superparticle size) $r_s$ by
\begin{equation} \label{eq:spreading} \Delta r(t) =\left(\frac{36 k N r_s^4 t}{t_{per}r}\right)^{1/3}, \end{equation}
where $N$ is the number of particles, $r$ is the radius of the ring, $t_{per}$ is the orbital period at that radius, and $k$ is a dimensionless constant. This equation provides a good fit to the simulation shown in Fig. (\ref{fig:spreading}) when we choose $k = 7.6\times10^{-4}$. With this constant, we can use equation (\ref{eq:spreading}) to choose a superparticle size that yields acceptable levels of numerical viscosity. For example, if we want to limit the spreading of the ring in this simulation to 10\% of its initial width in $10^7$ yr, we must choose a superparticle size less than $r_s = 10^{-4.12}$ AU.

\subsection{Conservation of Total Angular Momentum}

Collisions between planetesimals produce fragments smaller than the smallest size bin tracked by the superparticles. The mass of these dust particles is effectively lost to the superparticles, and carries away some angular momentum. We measured the angular momentum of the superparticles and the mass lost to small fragments in the planet-less ring simulation described in Section \ref{sec:heating} to verify that these torques balanced and that total angular momentum was conserved by SMACK. 

We calculated the time derivative of the angular momentum of the superparticles at each output timestep $t$ as
\begin{equation}\label{eq:dLdt} \left. \frac{dL}{dt} \right|_{superparticle} \approx \frac{\displaystyle\sum\limits_{i} L_i (t+\Delta t) - \displaystyle\sum\limits_{i} L_i(t)}{\Delta t}, \end{equation}
where $\Delta t$ is the size of an output timestep and $L_i(t)$ is the magnitude of the angular momentum of superparticle $i$ at time $t$, given by
\begin{equation}\label{eq:angmom} L_i(t) = m_i(t) |\textbf{v}_i(t)| |\textbf{r}_i(t)| \sin \theta_i(t), \end{equation}
where $\textbf{r}_i(t)$ and $\textbf{v}_i(t)$ are the the position and velocity vectors (in the heliocentric frame) of superparticle $i$, $m_i(t)$ is the total mass of superparticle $i$, and $\theta_i(t)$ is the angle between $\textbf{r}_i(t)$ and $\textbf{v}_i(t)$. 

We then set SMACK to output the mass lost in every superparticle encounter and the velocity of that mass. We calculated the time derivative of the angular momentum of the small fragments lost as dust in the same way as we calculated the time derivative of the superparticle angular momentum:

\begin{equation} \left. \frac{d\mathcal{L}}{dt} \right|_{dust} \approx \frac{\displaystyle\sum\limits_{j} \mathcal{L}_j(t+\Delta t) - \displaystyle\sum\limits_{k} \mathcal{L}_k(t)}{\Delta t}, \end{equation}
where $\mathcal{L}_j(t)$ is the angular momentum of the small fragments produced in each of the $j$ collisions during the output timestep $t$, and $\mathcal{L}_k(t)$ the fragments produced in the $k$ collisions during the previous output timestep.

\begin{figure}[!ht]
	\centering
	\includegraphics[scale=0.4]{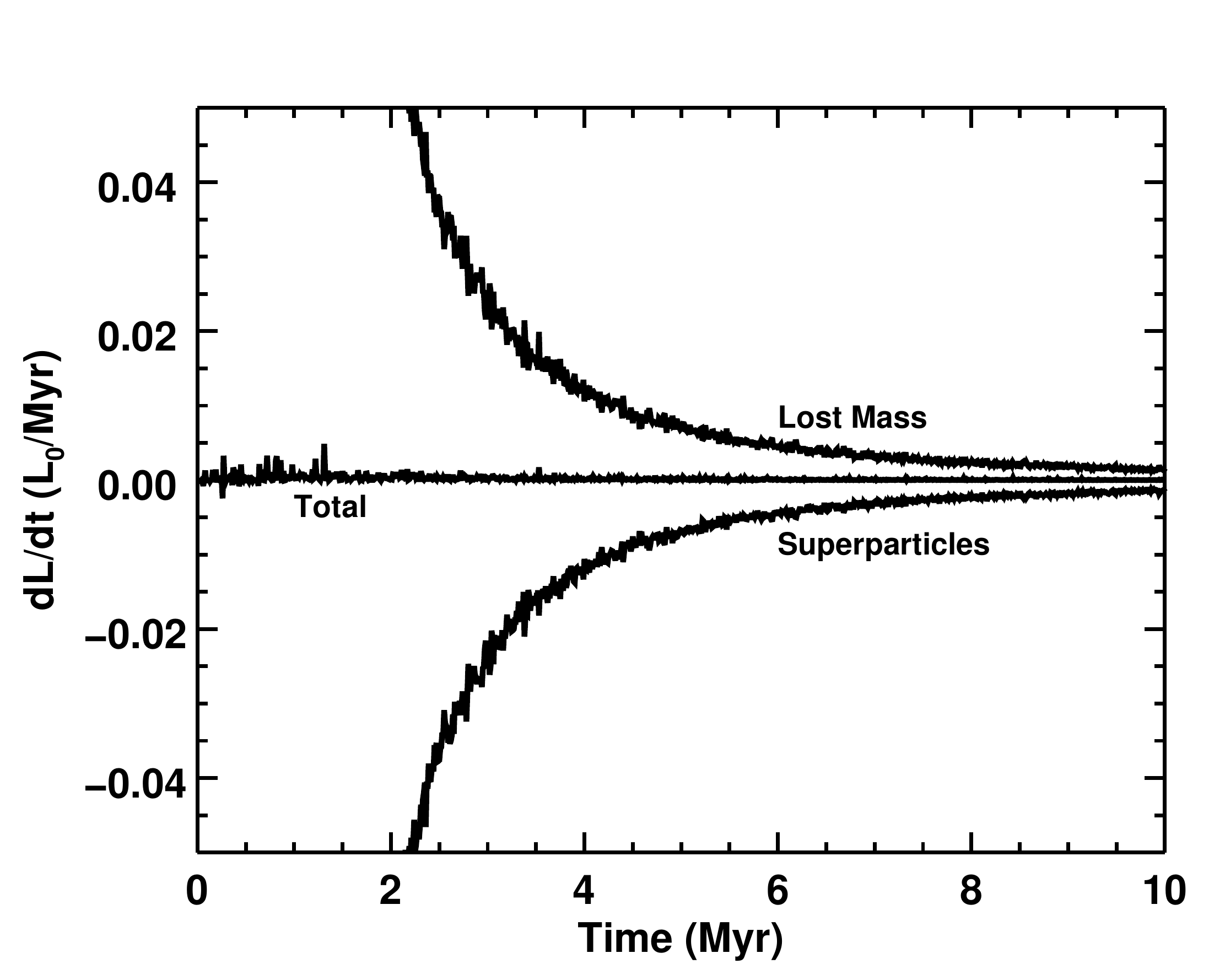}
	\caption{The time-derivative of the angular momentum $dL/dt$ of the disk over 10 Myr due to the superparticles and the mass lost to smaller fragments. The middle line shows the sum of $dL/dt$ for the superparticles and lost mass. The total angular momentum decay rate fluctuates stochastically around zero, but the system lost only 0.179\% of its initial angular momentum over the entire 10 Myr simulation to numerical noise.}
	\label{fig:dLdt}
\end{figure}

In Fig. \ref{fig:dLdt} we plotted $dL/dt$ of the superparticles, $d\mathcal{L}/dt$ of the lost mass, and the total for the two populations. The figure shows that most of the angular momentum lost by the superparticles is gained by the lost mass. The total change in angular momentum for the system varies stochastically over the simulation with a maximum variation of 0.48\% of the initial total angular momentum, $L_0$, per Myr. The systematic change in angular momentum is only $1.79\times10^{-3}$  of $L_0$ over 10 Myr. We also calculated the change in the x, y, and z-components of the system's angular momentum and found that the systemic change of each component was $-3.13\times10^{-3} L_0$,  $-4.46\times10^{-2} L_0$, and $1.79\times10^{-3} L_0$ over 10 Myr, respectively.

\subsection{Mass Loss Rates}

In the absence of velocity evolution, the rate of mass loss in a debris disk from planetesimal collisions is given by
\begin{equation} \dot{M}_{disk}(t) = -CM_{disk}^{2}, \end{equation}
and the total mass in a debris disk at time $t$ is given by
\begin{equation}\label{eq:massloss} M_{disk}(t) = \frac{M_0}{1+t/\tau}, \end{equation}
where $M_0$ is the initial mass of the disk, $\tau$ is a characteristic time at which  the disk mass is half its initial mass, and $C = 1/M_0 \tau$ \citep{Dominik2003}.

\begin{figure}[!ht]
	\centering
	\includegraphics[scale=0.4]{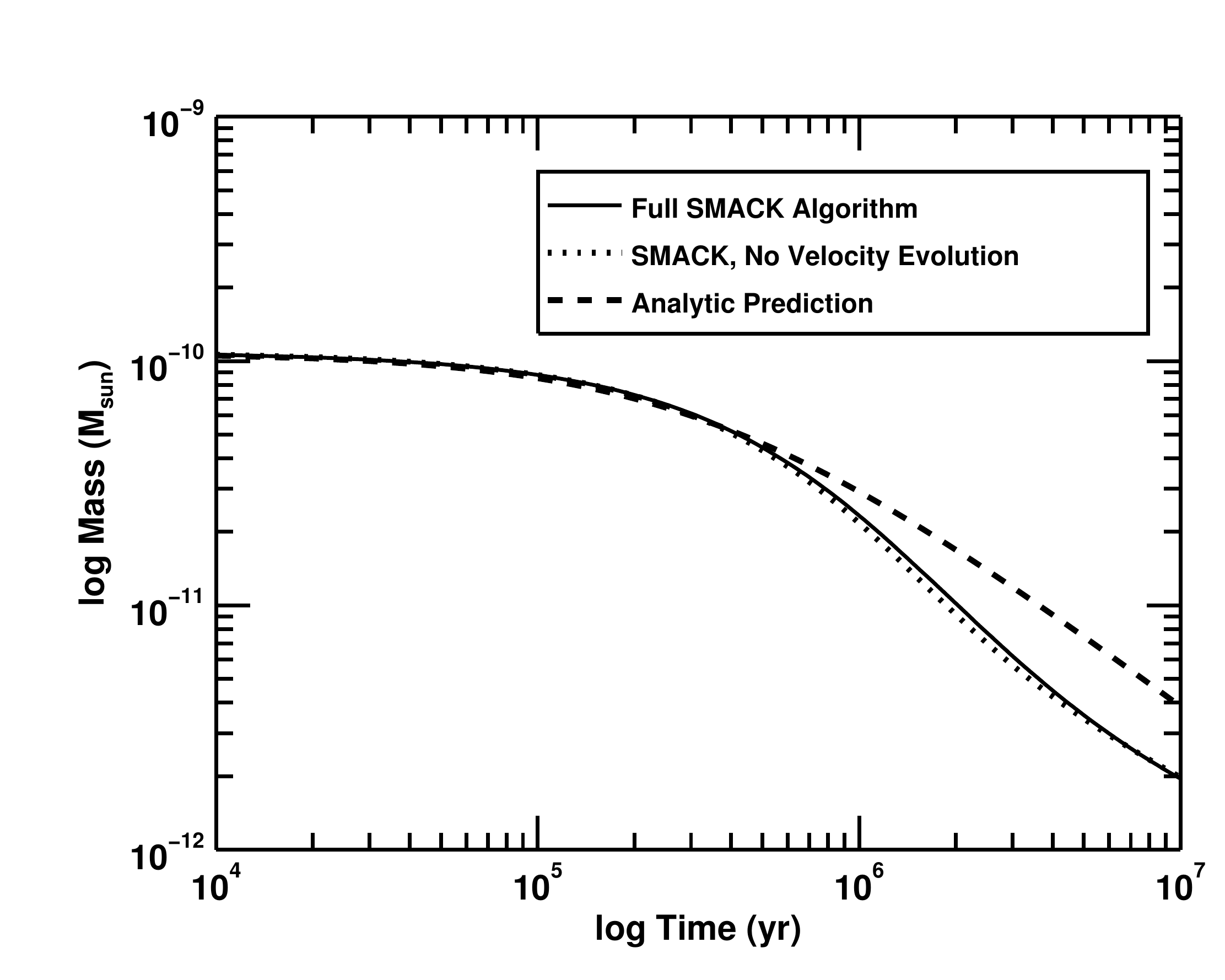}
	\caption{Total disk mass for a disk at $a=100$ AU with width $\Delta a = 20$ AU and initial maximum eccentricity $e = 0.2$ for a simulation without velocity evolution (dotted line) and with the full SMACK model  (solid line). Both simulations are in good agreement with the \cite{Dominik2003} analytic prediction of $M_{disk}(t) = M_0/(1+t/\tau)$ (dashed line) for $t<\tau = 1.4\times10^5$ yr, but the mass loss curves begin to diverge from from the analytic prediction for $t>\tau$. }
	\label{fig:massloss}
\end{figure}

We compared the evolution of the ring described in Section \ref{sec:heating} to this analytic expression. Fig (\ref{fig:massloss}) shows the total mass of the disk during each simulation, with the full SMACK model and also for an identical model with velocity evolution turned off (i.e., the velocities of the superparticles were not updated). The mass loss curves for each simulation are in good agreement with the \cite{Dominik2003} prediction for $t<\tau= 3.7\times10^5$ yr, but the simulations lose mass at a faster rate than predicted for $t>\tau$. The mass loss curve for the full SMACK simulation begins to flatten at $t\approx10^6$ yr as the mean eccentricity of the superparticles is damped (see Fig. (\ref{fig:heating})), decreasing the rate of fragmentation and slowing the mass loss. \cite{Lohne2008} also studied this problem with a model that included velocity evolution, but they did not report on this effect. More recently, \cite{Gaspar2012} used a collisional model to demonstrate a similar mass loss damping to the damping seen in our simulations in Fig. (\ref{fig:massloss}).

Rings of identical initial mass will have different characteristic times depending on their radii and the mean eccentricities of the planetesimals. \cite{Wyatt2007} analytically derived the following power law dependence of $C$ on the radius of a ring:
\begin{equation} C \propto r^{-13/3}. \end{equation}
This power law is the product of three contributions: the $r^{-1/2}$ dependence of the relative velocities in the ring, the $r^{-5/6}$ dependence of the total number of projectiles above the minimum required disruptive size on the impact velocities, and an $r^{-3}$ dependence of the density in the ring \citep{Lohne2008}. \cite{Lohne2008} studied mass loss rates in debris disks and replicated this $r^{-13/3}$ dependence in their simulations. In our simulations, the initial conditions of each system are determined by a user-input optical depth, so the density of the ring is independent of its size. This choice eliminates the $r^{-3}$ factor for the ring density, so we predict a power law dependence on ring radius of:
\begin{equation} C \propto r^{-4/3}. \end{equation}

\begin{figure}[!ht]
	\centering
	\includegraphics[scale=0.35]{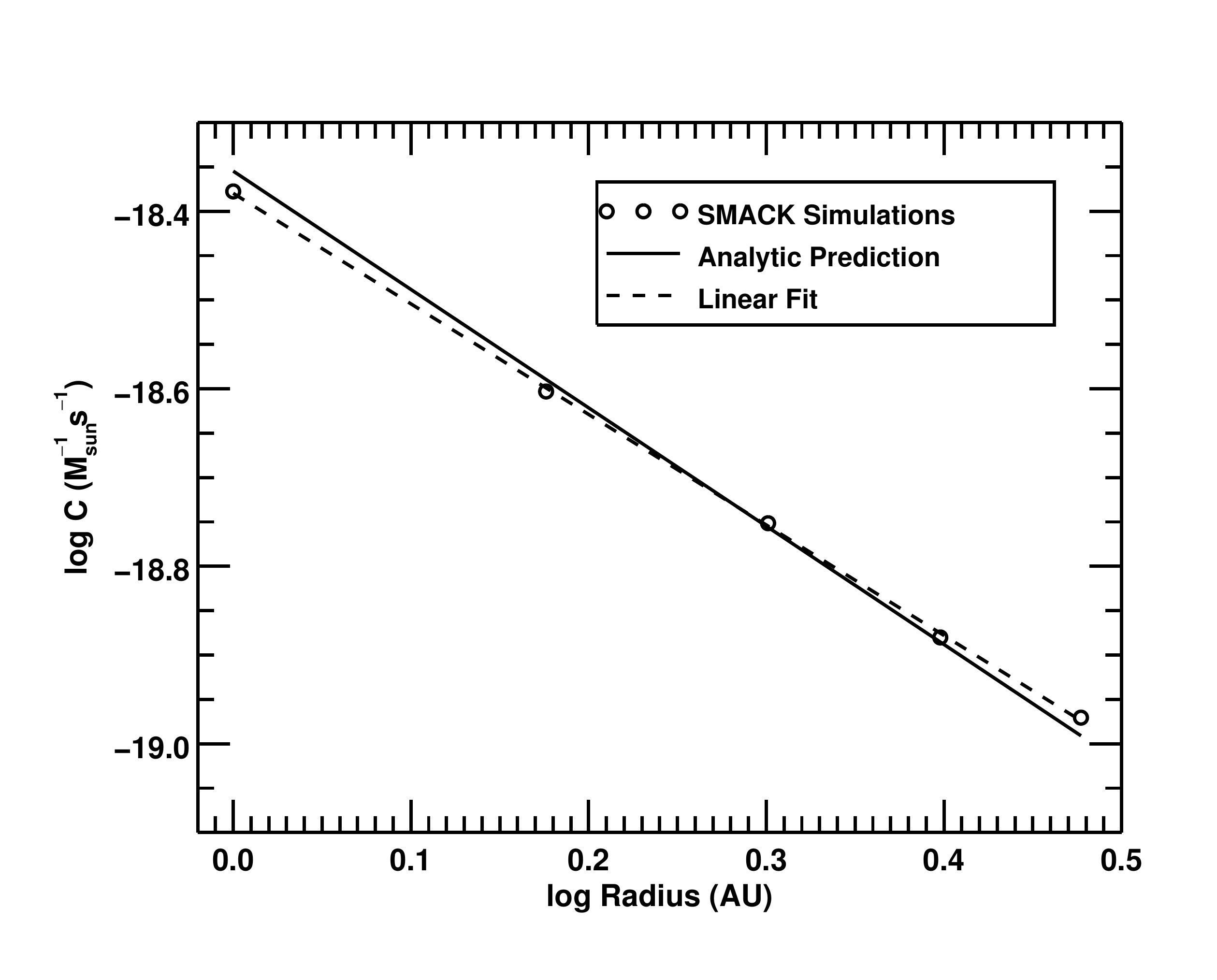}
	\caption{$C = 1/M_0 \tau$ vs. disk radius for a disk with eccentricity $e = 0.1$, width $\Delta a = 1$ AU, and no velocity evolution. The linear fit yields a power law with index -1.24542, close to the analytic prediction of $-4/3$.}
	\label{fig:CvsR}
\end{figure}

We measured the mass loss in three simulated rings of radial width $\Delta r = 1$ AU at five different radii from $r=1$ to $3$ AU, evolved for $10^4$ yr. We turned off SMACK's velocity evolution for these simulations by allowing SMACK to update the size distributions of the superparticles in each encounter without changing their velocities. We calculated $C$ for each ring and fit a power law to the relationship between $C$ and $r$ for each ring (Fig. \ref{fig:CvsR}). The resulting dependence of $C$ on radius in our simulated rings was $C \propto r^{-1.24542}$, in good agreement with the prediction of $C \propto r^{-4/3}$.

\cite{Wyatt2007} also derived the dependence of $C$ on the mean eccentricity, $e$, of the ring planetesimals:
\begin{equation} C \propto e^{5/3}. \end{equation}
\cite{Lohne2008} also tested this mass loss rate dependence on eccentricity, but found a power law dependence of $C \propto e^{9/4}$. They argued that the analytical model must be incomplete.

\begin{figure}[!ht]
	\centering
	\includegraphics[scale=0.35]{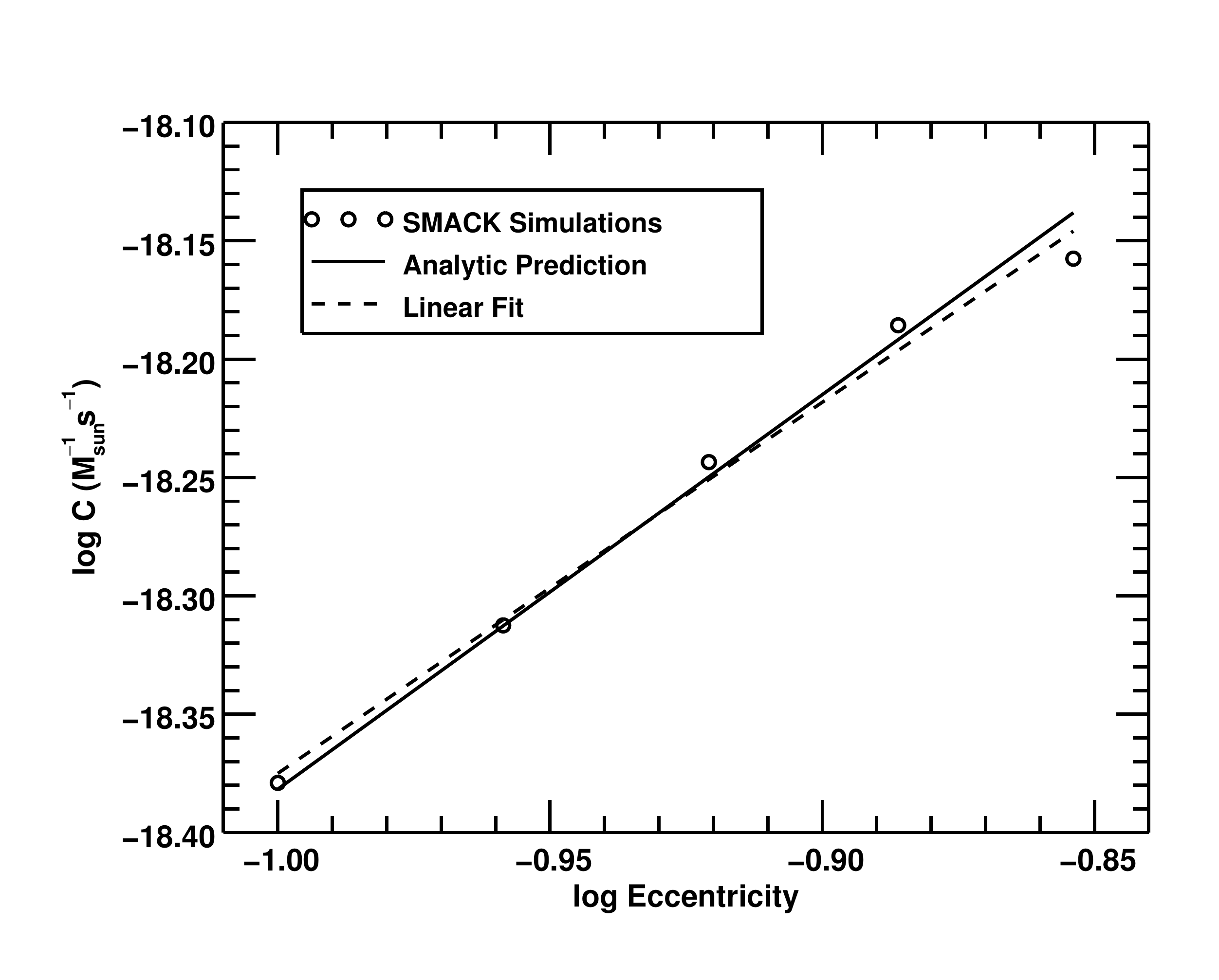}
	\caption{Disk eccentricity vs. $C = 1/M_0 \tau$ for a disk with $a = 1$ AU, width $\Delta a = 1$ AU, and no velocity evolution. The linear fit yields a power law with index 1.56767, close to the analytic prediction of $5/3$.} 
	\label{fig:CvsE}
\end{figure}

We measured the mass loss-eccentricity relationship in our models by simulating three rings at the same radius ($r=1$ AU) with different mean planetesimal eccentricities. Again, we turned off the velocity evolution in SMACK to compare the simulations with the predictions of \cite{Wyatt2007}. After measuring the mass loss in each ring, we calculated $C$ for each ring and fit a power law to the results (Fig. \ref{fig:CvsE}). The fit to the simulated data yields an index of 1.56767, in good agreement with the derived index of $5/3$. However, at higher mean eccentricities, the radial excursions of the particles on their eccentric orbits can begin to dominate the width of the ring. At this point, increasing eccentricity will decrease the optical depth of the ring, which slows the mass loss rate dependence on eccentricity and flattens out the $C$ vs. $e$ relationship. Fig. (\ref{fig:CvsE}) begins to show this effect in our simulations at higher eccentricities. 

\section{A Disk Containing a Planet on an Eccentric Orbit}
\label{sec:application}

To illustrate the application of SMACK to an interesting astrophysical system, we simulated the evolution of a debris ring containing a planet on an eccentric orbit. We simulated a disk of planetesimals using 10,000 superparticles with initial semi-major axes ranging from 90-110 AU around a solar-mass star. We used a superparticle radius of $r_s = 10^{-4/3}$ AU to maintain the same superparticle encounter time as the simulations in Fig. (\ref{fig:heating}). Using equation (\ref{eq:spreading}) we can see that with this superparticle radius, numerical viscosity will cause the ring to spread by less than a factor of $10^{-4}$ in 10 Myr.

The superparticle eccentricities and inclinations were uniformly distributed between $0-0.2$ and $0-0.1$, respectively. The remaining superparticle orbital elements $\Omega$, $\omega$, and $M$, were distributed uniformly between $0$ and $2\pi$. We inserted a planet with mass $8 M_{Jup}$ at 25 AU with eccentricity 0.5 and zero inclination. The vertical optical depth of the planetesimals at 100 AU from the star, $\tau_{disk}$, was set to $1 \times 10^{-2}$. We evolved the system for 10 Myr using SMACK. The simulation required only 4 wall clock hours on 48 cores on the NCCS Discover cluster. 

\begin{figure*}[!ht]
	\centering
	\includegraphics[scale=0.48]{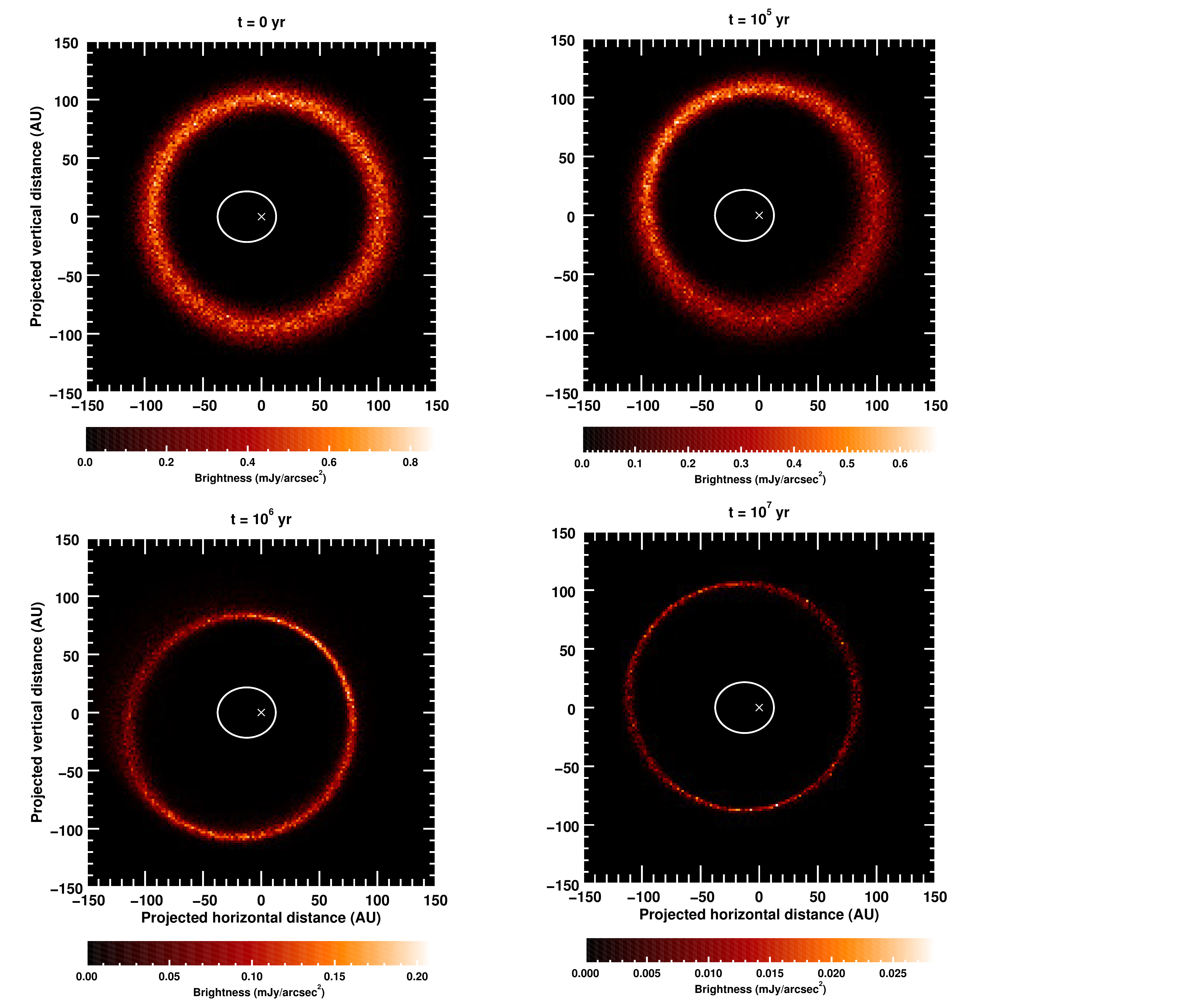}
	\caption{Simulated $850 \mu m$ images of  a disk with an $8 M_{Jup}$ planet with semi-major axis 25 AU and eccentricity 0.5 at $0$ yr, $10^5$ yr, $10^6$ yr, and $10^7$ yr. The white ellipse shows the planet's orbit; the white x indicates the location of the star. The maximum optical depth of the ring at $10^7$ yr (lower right frame) is $2.4\times10^{-4}$.}
	\label{fig:eccring}
\end{figure*}

SMACK output the coordinates and size distributions of the superparticles every $10^4$ yr. For each of four times during the simulation ($t = 0$, $10^5$, $10^6$, and $10^7$ yr), we used these coordinates and size distributions to synthesize images of the disk. To synthesize each image, we combined together the superparticle data from 10 output timesteps to reduce the Poisson noise in the image in the image, and projected the superparticles onto a 2D grid with resolution 2 AU to yield a face-on image. We then calculated each superparticle's contribution to the surface brightness in its pixel assuming each component planetesimal emits thermally as a spherical blackbody with
\begin{equation} I_{\nu SP} = f_{SP} \tau_{SP} B_\nu(T_{SP}), \end{equation}
where $B_\nu(T_{SP})$ is the value of the Planck function at $T_{SP}$, the temperature at the superparticle's distance from the star. Then we summed over all the superparticles in each pixel to find the total surface brightness of that pixel, assuming a central star with solar luminosity. We simulated images at a wavelength of $850 \mu$m. The results of the are shown in Fig. (\ref{fig:eccring}). 

The first image in Fig. (\ref{fig:eccring}), in the upper left, shows the initial conditions of the ring when the planet is added. The ring is circular and centered on the star. In the next image at $t=10^5$ yr, in the upper right, the planetesimal orbits are beginning to precess about the the planet's forced eccentricity. At $10^5$ yr the maximum vertical optical depth has dropped to $\tau_{disk} = 6.4\times10^{-3}$.
Note that the ring is brightest at the top left corner of this frame, not at periapse to the right of the frame. This increased surface brightness in the top left corner probably arises from differential precession; planetesimals with smaller semi-major axes precess slightly faster than those in the outer part of the ring. 

\begin{figure}[!ht]
	\centering
	\includegraphics[scale=0.43]{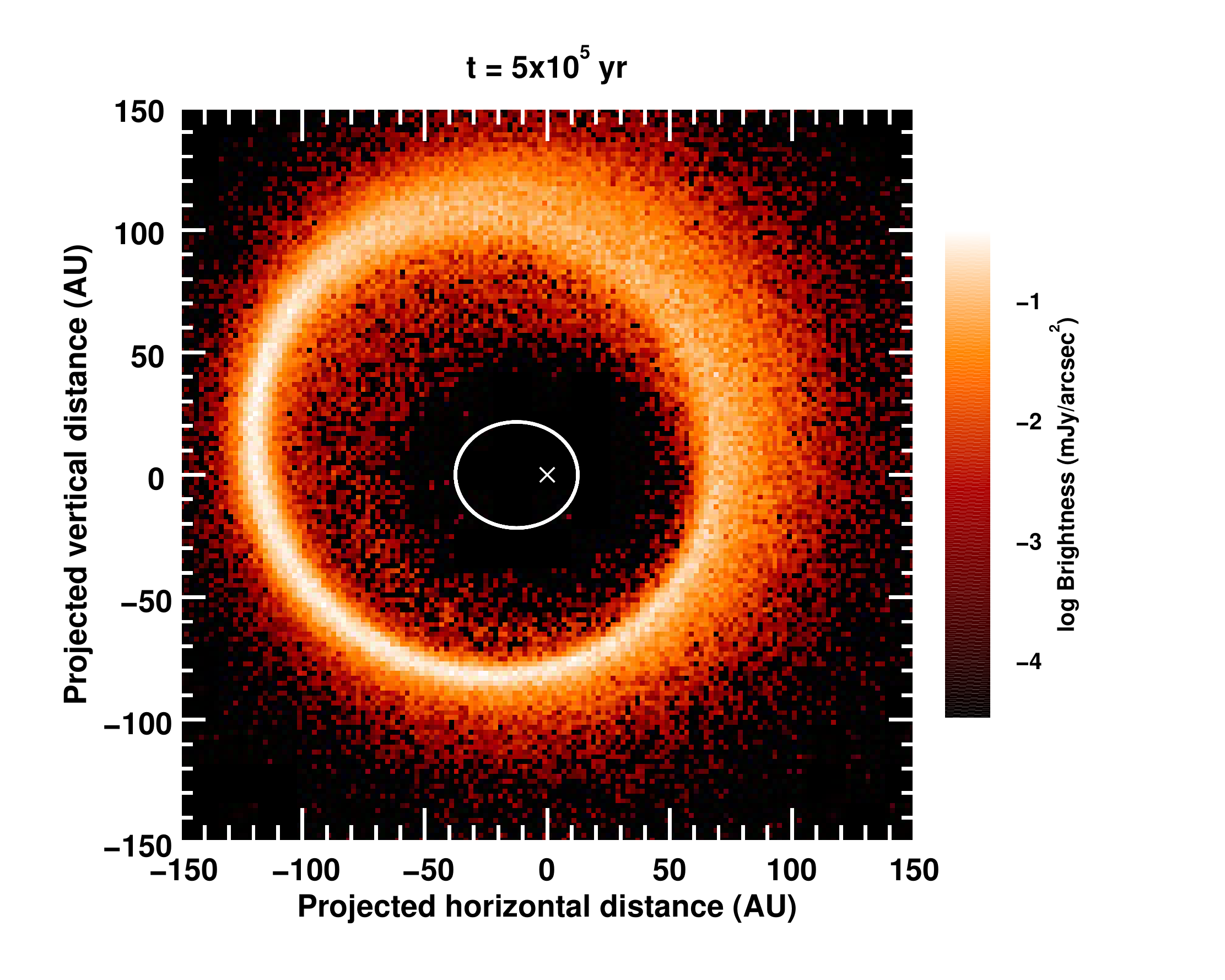}
	\caption{Simulated $850 \mu m$ image of  a disk with an $8 M_{Jup}$ planet with semi-major axis 25 AU and eccentricity 0.5 at $5\times10^5$ yr, from the same simulation shown in Fig. (\ref{fig:eccring}). The white ellipse shows the planet's orbit; the white x indicates the location of the star. The brightness scale in this image is logarithmic. The image shows a hint of spiral structure developing, before it is destroyed by collisions.}
	\label{fig:spiral}
\end{figure}

By $t = 10^6$ yr (lower left frame), differential precession has sheared the disk out into a hint of a spiral structure, as described in \cite{Wyatt1999}. This spiral can be seen most clearly in Fig. (\ref{fig:spiral}), which uses a logarithmic brightness scale for a simulated image $t = 5\times10^5 yr$. Then --- in stark contrast to \cite{Wyatt1999}, which did not incorporate collisions --- collisions break up the spiral before it can wrap once more around the star, leaving an apse-aligned ring shown in the lower right, as described by \cite{Quillen2006}. The maximum vertical optical depth at t = $10^6$ yr is $\tau_{disk} = 1.7\times10^{-3}$.
The apse-aligned ring that remains at $t = 10^7$ yr (lower right frame) is now narrower, with a width of $\sim 15$ AU, because collisions remove objects in orbits that are not approximately nested after precessing differentially for the lifetime of the system. At this final stage, the maximum vertical optical depth in the ring has dropped to $\tau_{disk} = 2.4\times10^{-4}$.

Inelastic collisions between particles can both excite and damp random velocities \citep{Goldreich1978}. Some authors have assumed that planetesimals in a debris disk have had their free eccentricities completely damped by collisions \citep{Quillen2006, Chiang2009} resulting in a ring of planetesimals on orbits with eccentricity equal to the forced eccentricity from the planet. Farther from the planet, the magnitude of the forced eccentricity is given by
\begin{equation} |e_{forced}| = \frac{b_{3/2}^2 (\alpha)}{b_{3/2}^1 (\alpha)} e_p, \end{equation}
where $e_p$ is the planet's eccentricity, $b_{3/2}^2 (\alpha)$ and $b_{3/2}^1 (\alpha)$ are Laplace coefficients, and $\alpha$ is the ratio of the planet's semi-major axis $a_p$ to the ring's semi-major axis $a$ for $a_p < a$ \citep{Murray1999}. In the disk shown in Fig. (\ref{fig:eccring}), the forced eccentricity at a semi-major axis of 100 AU is $\sim 0.155$. The mean eccentricity of the planetesimals in the $t=10^7$ image in Fig. (\ref{fig:eccring}) is $\sim 0.224$, indicating that the free eccentricities have not been completely damped, and the ring is not yet collisionally relaxed.

Besides secular perturbations and collisions, mean motion resonances can also sculpt debris rings, clearing central holes \citep[e.g.][]{Mustill2012} and creating a variety of azimuthal structures \citep[e.g.][]{Kuchner2003}. However, the inner edge of the ring in our simulation is far outside the region of resonance overlap and an inspection of the time-averaged semi-major axes of the surviving superparticles near the end of the simulation ($10^7$ yr) suggests that only about 65 out of 1500 occupy the strongest mean motion resonances with the planet (7:1, 8:1, and 9:1) that intersect the ring. More simulations will be needed to study the role of resonances in collisional rings, and the ability of SMACK to model such fine phase space structures using finite-sized superparticles. But for now, it appears that mean motion resonances play no more than a minor role in the evolution of this system, and that secular and collisional evolution dominate the dynamics.

To check if the erosion of the ring edges could be a numerical artifact, we ran the same simulation with four times as many superparticles. A histogram of the final semi-major axis distribution for the higher-fidelity simulation was nearly identical to final semi-major axis distribution of the simulation shown in Fig. (\ref{fig:eccring}), with differences consistent with Poisson noise. This comparison suggests that the narrowing of the ring illustrated in the lower right of Fig. (\ref{fig:eccring}) is real, caused by the combined effects of collisional damping and collisional (viscous) stirring of the differentially-precessing ring.

\begin{table*}
\centering
\begin{tabular}{c c c c c c c c}
Figure & $a$ (AU) & $e_{max}$ & $i_{max}$ & $\tau$ & $l_{box}$ (AU) & $r_s (AU)$ & Velocity Evolution\\
\tableline
(\ref{fig:slopes}) 	& 90-110			& 0.2 	& 0.1  	& $1\times10^{-2}$ 	& 390	&$10^{-1}$& Off\footnote{Indicates that SMACK did not update the velocities of the superparticles due to collisions.}\\
(\ref{fig:heating})	& 90-110    		& 0.2	 	& 0.1  	& $1\times10^{-2}$ 	& 390	&Varies\footnote{Indicates that multiple simulations were run with different values of the given parameter.}     & On\footnote{ Indicates that SMACK updated the superparticle velocities using Equations (\ref{eq:va}) and (\ref{eq:vb}).}\\
(\ref{fig:spreading})	& 0.995-10.005		& 0.1      	& 0.05      	& $1\times10^{-3}$	& 30		&$10^{-2.5}$& On\\
(\ref{fig:dLdt})	        	& 90-110            	& 0.2		& 0.1	 	& $1\times10^{-2}$ 	& 390	&$10^{-1.3}$& On\\
(\ref{fig:massloss})	& 90-110			& 0.2		& 0.1		& $1\times10^{-2}$ 	& 390	&$10^{-1.3}$& Both\footnote{Indicates that SMACK updated the superparticles in some of the simulations but not others. }\\
(\ref{fig:CvsR})	        	& Varies			& 0.1 	& 0.0    	& $1\times10^{-3}$ 	& 10		&$10^{-3}$& Off\\
(\ref{fig:CvsE})	 	& 0.5-1.5            	& Varies	& 0.0    	& $1\times10^{-3}$ 	& 10		&$10^{-3}$& Off\\
(\ref{fig:eccring})	& 90-110            	& 0.2		& 0.1		& $1\times10^{-2}$ 	& 390	&$10^{-4/3}$& On\\
\end{tabular}
\caption{Initial conditions for all simulations in Sections \ref{sec:tests} and \ref{sec:application}. }
\label{tab:initial}
\end{table*}

\section{Current Limitations of SMACK}
\label{sec:limitations}

The main limitations of the current version of SMACK derive from using the superparticles to each represent many decades in planetesimal mass. Since the dominant size bins tend to be the smallest size bins and the ``swapping'' described in Section \ref{sec:smack} tends to mix the size distributions among the superparticles, the current version of the code tends to make errors in the position and velocity distributions of larger bodies. For example, the current version of SMACK does not accurately model mass segregation and the velocity distributions of planetesimals of different sizes, as in the work of \cite{Pan2011}. 

To investigate the level of mass segregation that SMACK can reproduce, we simulated the evolution of a planetless ring of superparticles with initial eccentricity of 0.1, as described in Section \ref{sec:heating}. We chose a superparticle radius of 0.1 AU to minimize numerical heating. We found that the eccentricity of the 1 mm planetesimals damped faster that the eccentricity of the 1 m planetesimals by roughly a factor of 2, i.e., the planetesimals in SMACK do undergo some mass segregation, but not enough.

In a real disk, the damping rate for planetesimals of size $D$ by bodies of size $s \leq D$ is 
\begin{equation} \frac{1}{\tau_{damp}} \propto \frac{N(s) s^3}{D}, \end{equation}
where $\tau_{damp}$ is the damping timescale for the planetesimals and $N(s)$ is the incremental size distribution of the disk with logarithmic size bins \citep{Pan2011}.
Assuming a \cite{Dohnanyi1969} size distribution of $N(s) \propto s^{-2.5}$, the damping rate goes as $s^{0.5} D^{-1}$. Collisions for which $s \approx D$ will dominate, so the collision rate is proportional to $D^{-0.5}$. We therefore expect the damping timescale for 1 m planetesimals to be larger than the timescale for 1 mm planetesimals by a factor of $(10^3)^{1/2} \approx 32$, substantially larger than our SMACK model. In future versions of SMACK, we will explore using more superparticles and limiting the range of planetesimal sizes that each superparticle represents to improve SMACK's ability to model mass segregation in disks.

SMACK is based on the following assumptions:
\begin{itemize}
\item{No radiative forces: Besides causing issues with mass segregation, the use of superparticles also requires that all planetesimals in a superparticle be subject to the same forces. For example, radiation pressure and Poynting-Robertson drag forces both vary with the size of the planetesimal. In this paper, we restrict our size distributions to planetesimals larger than 1 mm, for which radiative forces are not significant in a typical debris disk.}
\item{No interaction within a superparticle: SMACK also assumes that the planetesimals within a superparticle do not interact. A cloud of planetesimals orbiting a star with the similar trajectories would not experience fragmenting collisions between planetesimals due to their low relative velocities, but they could still experience collisions with other outcomes, such as bouncing or sticking. Also, larger planetesimals in the cloud would gravitationally influence the other bodies. SMACK does not simulate gravity between planetesimals, so it cannot model grain growth or accretion by gravity.}
\item{No gravitational interaction between superparticles: Aside from preventing grain growth, the lack of gravitational interaction between large bodies in SMACK also prevents modeling of dynamical effects such as viscous stirring.}
\item{Simplified crushing law: The crushing law used in this version of SMACK models only catastrophic collisions, with no sticking, cratering, or rebounding collisions. SMACK also assumes a single, mass-independent material strength, and calculates the size of the largest fragment independent of impact velocity. This simplified crushing law can reproduce a size distribution evolution to a basic \cite{Dohnanyi1969} equiliubrium. Future versions of SMACK will include more complex crushing laws based on experimental results and SPH modeling to investigate the effects of crushing law parameters on the size distribution evolution of a disk.}
\end{itemize}

\section{Summary and Future Work}
\label{sec:future}

SMACK, a new collisional module for the REBOUND N-body integrator, uses a superparticle method to simulate the evolution of the dynamics and size distribution of a debris disk experiencing fragmenting collisions. SMACK models the interaction between the disk and embedded planets in 3-D for planetesimals with a range of sizes and can model various disk asymmetries including eccentric rings and warps. When run on parallel CPUs, SMACK can simulate the evolution of disks older than $10^7$ yr in a feasible number of hours.

We showed that SMACK, with its simple ``swapping'' algorithm, can reproduce a variety of fundamental results in debris disk physics. With the velocity evolution turned off (i.e., by not updating superparticle velocities), we used SMACK to reproduce the evolution of the size distribution of planetesimals to a power-law with index $-2.5$, using incremental logarithmic size bins, as \cite{Dohnanyi1969} derived analytically. We verified that SMACK conserves total angular momentum during dust-producing collisions and showed that we could mitigate the effects of numerical heating and numerical viscosity by constraining the size of the superparticles. We reproduced the mass loss function for a debris disk derived by \cite{Dominik2003} and demonstrated that the rate of mass loss varies with disk radius and planetesimals mean eccentricity as roughly $r^{-4/3}$ and $e^{5/3}$, as \cite{Wyatt2008} derived. 

After performing this battery of tests, we used SMACK to simulate the evolution of a disk containing a central planet on an eccentric orbit (Figure \ref{fig:eccring}). The simulation shows how planetesimal collisions, together with the planet's gravity, converted a circular ring of planetesimals (width 20 AU), centered on the star, into a narrower (6 AU) ring, offset from the star. It illustrates the combined effects of secular precession, collisional damping and collisional (viscous) stirring. Previous models of a planet on an eccentric orbit interacting with a debris disk have assumed either no particular collisional evolution of the planetesimal orbits \citep{Wyatt1999, Wilner2002, Quillen2002, Moran2004}, or complete collisional damping \citep{Quillen2006, Chiang2009}. Our model hints at the wide range of intermediate possibilities that are likely important in observed debris disks; in our simulation, the planetesimals were still not completely collisionally damped after $10^7$ yr, roughly the age of $\beta$ Pictoris. 

\cite{Boley2012} have suggested that a pair of planets might be needed to create the narrow planetesimal ring around Fomalhaut. But no second planet was required in our simulation. Of course, the scenario we simulated was ultimately an artificial one: a planet on an eccentric orbit suddenly introduced into a disk of planetesimals in a circular ring. More simulations, with a wider range of initial conditions and planet-formation scenarios are needed to unravel the physics of the Fomalhaut system.

The primary limitation of the SMACK models presented in this paper at that they underestimate the mass segregation rate, as described in section (\ref{sec:limitations}). We plan to try improving the way SMACK models planetesimal-size-dependent effects by limiting the range of planetesimal masses that each superparticle represents. We hope to extend the large end of the size distributions to 1 km and beyond.

We anticipate several other future improvements. The current version of SMACK uses a size-independent breaking strength for the planetesimals, which we will replace with a size-dependent strength to explore its effects on the steady-state size distribution of a collisional cascade. We will also expand the set of collisional outcomes to include cratering and bouncing collisions. Aggregation and planetesimals growth could also conceivably be simulated with SMACK.

We also intend to implement a dynamic and adaptive domain decompositioning for REBOUND to enable us to run simulations on even more CPUs. This will allow us to improve the running time of REBOUND and SMACK and to model even larger systems.

With the flexibility of the REBOUND platform, SMACK can model a wide range of debris disk scenarios: disks with planets on eccentric orbits (e.g. Section \ref{sec:application}), disks with planets on inclined orbits like the $\beta$ Pictoris system \citep{Lagrange2009, Lagrange2010}, even multiple planet systems and migrating planets. Millimeter and sub-millimeter observations with ALMA are beginning to probe populations of larger dust grains and parent bodies in debris disks at high angular resolution \citep[e.g.][]{Boley2012, MacGregor2013}. We hope to make SMACK a workhorse for interpreting ALMA images of debris disks.

We thank the NASA High-end Computing Program for granting us time on the Discover cluster. This research was supported in part by  NASA Planetary Geology and Geophysics Program, grant no. 11-PGG11-0032.

\bibliographystyle{apj}
\bibliography{libraryv2}

\newpage

\end{document}